\documentclass[12pt,a4paper,preprint,superscriptaddress]{revtex4-1} 

\usepackage{float}
\usepackage{amsmath,amssymb}
\usepackage[utf8]{inputenc}
\usepackage[english]{babel}
\makeatletter\AtBeginDocument{\let\@elt\relax}\makeatother
\usepackage{graphicx}

\begin{document}


\title{Generalized Method for Charge Transfer Equilibration in Reactive Molecular Dynamics}

\date{\today}
\author{Tobias Gergs}
\affiliation{Department of Electrical Engineering and Information Science, Ruhr University Bochum, 44801 Bochum, Germany}
\affiliation{Electrodynamics and Physical Electronics Group, Brandenburg University of Technology Cottbus-Senftenberg, Siemens-Halske-Ring 14, 03046 Cottbus, Germany}
\author{Frederik Schmidt}
\affiliation{Department of Electrical Engineering and Information Science, Ruhr University Bochum, 44801 Bochum, Germany}
\author{Thomas Mussenbrock}
\affiliation{Department of Electrical Engineering and Information Science, Ruhr University Bochum, 44801 Bochum, Germany}
\author{Jan Trieschmann}
\affiliation{Electrodynamics and Physical Electronics Group, Brandenburg University of Technology Cottbus-Senftenberg, Siemens-Halske-Ring 14, 03046 Cottbus, Germany}

\begin{abstract}
Variable charge models (e.g., EEM, QEq, ES+) in reactive molecular dynamics simulations often inherently impose a global charge transfer between atoms (approximating each system as ideal metal). Consequently, most surface processes (e.g., adsorption, desorption, deposition, sputtering) are affected, potentially causing dubious dynamics. This issue is meant to be addressed by the ACKS2 and QTPIE model, which are based on the Kohn-Sham density functional theory as well as a charge transfer restricting extension to the QEq model (approximating each system as ideal insulator), respectively. In a brief review of the QEq and the QTPIE model, their applicability for studying surface interactions is assessed in this work. Following this reasoning, the demand for a revised generalization of the QEq and QTPIE model is proposed, called charge transfer equilibration model or in short QTE model. This method is derived from the equilibration of constrained charge transfer variables, instead of considering atomic charge variables. The latter, however, are obtained by a respective transformation, employing an extended Lagrangian method. We moreover propose a mirror boundary condition and its implementation to accelerate surface investigations. The models proposed in this work facilitate reactive molecular dynamics simulations which describe various materials and surface phenomena appropriately.
\end{abstract}

\maketitle

\newpage


\section{Introduction}
\label{sec:introduction}

Molecular dynamics can be subdivided into three groups, i.e., classical, reactive and ab initio molecular dynamics. The first is commonly used to investigate simple processes on larger length and time scales. The last allows for a thorough study of complex material compositions and dynamics at the cost of substantially higher computational resources. The gap between these two methods is addressed by reactive molecular dynamics (RMD), which typically employs the concept of bond order in combination with variable (or fluctuating) charge methods. The bond order of an atom is used to describe its environment-dependent interatomic bond strength. Variable charge models allow for a corresponding environment-dependent charge distribution. The latter is determined by fulfilling Sanderson's electronegativity equalization (EE) within the system \citep{sanderson1951interpretation}. This corresponds to the minimization of the overall electrostatic energy under the constraint of charge neutrality and a fixed atom geometry \citep{parr1978electronegativity}. Some of the most prominent self-consistent variable charge models are the electronegativity equalization method (EEM) \citep{mortier1986electronegativity, mortier1985electronegativity, bultinck2002electronegativity1, bultinck2002electronegativity2}, charge equilibration (QEq) \citep{rappe1991charge} and electrostatic plus (ES+) \citep{streitz1994electrostatic}. Apart from parameter definitions, they differ from each other in the way the atomic charge is spatially distributed (in QEq, the hydrogen electronegativity is also meant to be charge-dependent). In EEM, point charges and eventually a shielded Coulomb potential are used to model electrostatic interactions \citep{njo1998extending}. In QEq, a single normalized $n$s Slater orbital is used to describe the outer valence orbital. Furthermore, the diatomic Coulomb integral is evaluated. ES+ extends the QEq model by additional consideration of the core charge. Hence, RMD potentials (e.g., ReaxFF \citep{van2001reaxff, senftle2016reaxff} and COMB3 \citep{liang2013classical}) that make use of both concepts (bond order and variable charge) enable simulations of particular complex material compositions or phenomena. 

The bidirectional transition from one system to two or more non-bonded systems (e.g., dissociation, recombination, desorption, adsorption, sputtering, deposition, fragmentation) as well as the interaction of the latter with each other (e.g., two or more distant molecules) cause issues when applying EEM, QEq or ES+. All models allow for a non-physical charge transfer between spatially separated atoms, molecules and solids \citep{senftle2016reaxff}. The electronegativity is equalized within the total system, without any geometrical limitations. This corresponds to a global charge transfer between all atoms until the EE is reached. Even a single system is therefore always approximated as an ideal conductor (metal) \citep{nistor2009dielectric, chen2009theory}. However, manifold RMD studies of insulators, where either EEM, QEq or ES+ was applied, have proven that at least this circumstance can be dealt with by the respective RMD formalism. The metallic approximation eventually causes issues though, when the polarization due to an external electric field is important (e.g., resistive switching mechanisms) \citep{gergs2018integration}.

This methodical challenge has been addressed for instance by applying an extended Lagrangian method. It was proposed to split up the system into subsystems (e.g., molecules) and solve for the EE only within those (allowing for intramolecular, but omitting intermolecular charge transfer) \citep{rick1994dynamical}. The drawback of this approach is the inherent inability of describing any bidirectional transition between those systems (e.g., dissociation, recombination) due to the fixed subsystem definitions. 

The split charge equilibration (SQE) method is based on EEM, but makes use of diatomic charge transfer instead of atomic charge variables \citep{mathieu2007split, krykunov2017new}. While this formalism enables a straightforward way to mitigate long range charge transfer and overall adjust the latter more precisely, it also increases substantially the number of charge-related variables and thus computational cost. The atom-condensed Kohn-Sham density functional theory approximated to second order (ACKS2) model is a generalization of SQE, however, employing atomic charge variables and being derived from the Kohn-Sham density functional theory\citep{verstraelen2013acks2, verstraelen2014direct}. Recently, it replaced EEM in the ReaxFF potential, but for example yet needs to be implemented in the widely used open-source molecular dynamics simulation framework LAMMPS \citep{senftle2016reaxff, plimpton1995fast}.

For the charge transfer polarization current equalization (QTPIE) method, initially, charge transfer variables are used. In addition, diatomic electronegativity differences (for neutral atoms) are scaled with the $n$s-type overlap integral to constrain the charge transfer spatially \citep{chen2007qtpie, chen2008unified, chen2009theory}. QTPIE is a generalization of QEq. Later, the charge transfer variables are transformed back to atom charge variables and as a final result, effective electronegativities are defined. These can be thought of as being a weighted average of the beforehand mentioned scaled electronegativity difference. Apart from that, QTPIE corresponds to QEq. In the frame of the QTPIE method, all systems are approximated as ideal insulators \citep{chen2009theory}. Thus, the QTPIE method is not suited for the simulation of metals or semiconductors.

The comparison of QEq with QTPIE leads to the representation via two models, which can be interpreted as describing the systems in two limiting cases. In QEq (EEM, ES+), any system is approximated as an ideal conductor (metal) with global charge transfer. In QTPIE, any system is approximated as an ideal insulator. Both models are reviewd in Section~\ref{sec:revision_of_variable_charge_models}. In Section~\ref{ssec:charge_transfer_equilibration}, a charge transfer equilibration (QTE) model is proposed that addresses the gap between these two limiting cases. QTE is therefore an alternative to ACKS2, which is based on a different approach. In addition, in Section~\ref{ssec:extension_to_QEq}, a minor extension to the still widely used EEM, QEq and ES+ is described, which allows for a better charge transfer in case of specific surface processes (i.e., adsorption, desorption). Subsequently, in Section~\ref{sec:mirror_boundary_condition}, a mirror boundary condition is described, which accelerates RMD simulations that employ variable charge models. In Section~\ref{sec:implementation}, recommendations for the respective implementations are provided. The models are validated in Section~\ref{sec:validation} by comparing them for a small set of demonstration cases. One of which is then used in Section~\ref{sec:performance} to estimate the charge models individual performance. Finally, in Section~\ref{sec:discussion}, a conclusion is presented.


\section{Review of variable charge models}
\label{sec:revision_of_variable_charge_models}

In the following, EEM \citep{mortier1986electronegativity, mortier1985electronegativity, bultinck2002electronegativity1, bultinck2002electronegativity2}, QEq \citep{rappe1991charge}, ES+ \citep{streitz1994electrostatic} and QTPIE \citep{chen2007qtpie, chen2008unified, chen2009theory} will be briefly summarized. All of which lead to a set of coupled linear equations, which can be solved in different ways. Here, an extended Lagrangian method for treating fictitious degrees of freedom (atomic charge space) was chosen, as proposed elsewhere \citep{rick1994dynamical}. On the one hand, this approach is utilized by a frequently used RMD potential (COMB3) \citep{liang2013classical}. On the other hand, we believe, that this method allows for a more intuitive interpretation of those models. ACKS2 and SQE are not revisited, since their approaches differ immensely and, therefore, do not provide further insight in this context.

\subsection{Extended Lagrangian method for EEM, QEq and ES+}
\label{ssec:extended_lagrangian_method_for_EEM_QEq_and_ES+}

Apart from parameter definitions, EEM, QEq and ES+ differ in the way the atomic net charge is spatially described. The respective electrostatic interaction $J_{ij}$ (hardness in case of $i=j$) between atom $i$ and atom $j$ is therefore different. The electronegativity $\chi_i^0$ for a neutral atom $i$ is, however, consistent throughout these models. Thus, as long as it is not necessary to specify $J_{ij}$, all models can be discussed at once.

The overall goal to determine atomic charge distribution $q_i$ is accomplished by the EE. The electronegativity of the $i$-th atom $\chi_i$ can be described by the negated chemical potential of the electrons $\mu_i$ surrounding their nucleus $i$,

\begin{subequations}
\begin{align}
\chi_i 		   &= - \mu_i = - \frac{\partial E}{ \partial N_\mathrm{e}} = e\frac{\partial E}{\partial q_i} 
\label{eq:electronegativity_definition1}
\end{align}
\end{subequations}

where $E$ is the total energy of the system and $N_\mathrm{e}$ is the number of electrons \citep{parr1978electronegativity}. A second-order Taylor expansion of the energy with respect to the atomic charge $q_i$ allows one to define the normalized electronegativites $\tilde{\chi}_i = \frac{1}{e}\chi_i$ and $\tilde{\chi}_i^0 = \frac{1}{e}\chi_i^0$ as

\begin{subequations}
\begin{align}
\tilde{\chi}_i &= \tilde{\chi}_i^0 + \sum_{j=1}^N J_{ij}q_j,
\label{eq:electronegativity_definition2}
\end{align}
\end{subequations}

where $N$ is the number of atoms \citep{parr1978electronegativity,rappe1991charge,iczkowski1961electronegativity,rappe1991charge}. 

The respective Lagrangian is defined by 

\begin{equation}
\label{eq:lagrangian}
\mathcal{L} = \sum_{i=1}^N \frac{1}{2}m_i \boldsymbol{\dot r_i}^2 + \sum_{i=1}^N \frac{1}{2}m_{q} {\dot q_i}^2 + - U[\{\boldsymbol q \},\{\boldsymbol r \}] - \lambda \sum_{i=1}^N q_i,
\end{equation}

where $U$ is the potential energy of the system, $\boldsymbol{r_i}$ is the nuclei site, $m_i$ is the atom mass, $m_{q}$ is the fictitious charge mass and $\lambda$ is the Lagrange multiplier. Indexed and plain bold letters indicate vectors and tensors of the complete system space, respectively. The Lagrange multiplier is meant to enforce charge neutrality $\sum_{i=1}^N q_i = 0$. 

In Lagrange mechanics, variables are meant to be independent from each other. Otherwise, constraints may be used to form a set of generalized coordinates. The degrees of freedom due to the atomic charges are introduced as fictitious coordinates, defining an equivalent charge coordinate space. The nuclei sites and the atomic charges can be approximated to be independent from each other. Since the electron dynamics are orders of magnitude faster than the nuclei dynamics, the charges are commonly assumed to be equilibrated for each nuclei movement. This means that the nuclei and the charge dynamics are solved sequentially and not in parallel. Thus, the nuclei are immobile during the charge dynamics and therefore a constant of the fictitious motion. However, for clarity, we will discuss this point briefly again during the following derivations, when devising the Euler-Lagrange equation.

The time evolution of the atomic charge distribution is described by  
\begin{subequations}
\begin{align}
m_{q} {\ddot q_i} &= - \lambda - \frac{\partial U}{\partial q_i} - \sum_{j=1}^N \nabla_j U \frac{\partial \boldsymbol r_j}{\partial q_i}
\label{eq:charge_motion1}\\
					  &= - \lambda - \tilde{\chi}_i ,
\label{eq:charge_motion2}
\end{align}
\end{subequations}

where \eqref{eq:electronegativity_definition1} is used and, as mentioned beforehand, the nuclei sites are constant during the fictitious charge motion and thus, the second subtrahend is zero. By summation over all atoms (from 1 to $N$), $\lambda$ can easily be found to be the negated average electronegativity $\tilde{\overline{\chi}} = \frac{1}{N}\sum_{j=1}^N \tilde{\chi}_j$. The final equation for the fictitious charge motion is
\begin{equation}
\label{eq:charge_motion_final}
m_{q} {\ddot q_i} = \tilde{\overline{\chi}} - \tilde{\chi}_i = \sum_{j=1}^N \frac{\tilde{\chi}_j - \tilde{\chi}_i}{N}.
\end{equation}

Due to the computational effort, the EE is even for the equilibrated system only fulfilled up to the set tolerance (precision), $\ddot q_i \approx 0~\forall i \in [1,N]$. Hence, the following equation holds for the chemical potential by applying equation \eqref{eq:electronegativity_definition1},
\begin{equation}
\label{eq:chemical_potential_final}
\mu_i = -\chi_i \approx -\overline{\chi} ~~\forall~i \in [1,N].
\end{equation}

In case of an external electric field $\boldsymbol{E} = -\nabla \Phi$, the equation for the fictitious charge motion become
\begin{equation}
\label{eq:charge_motion_external_field}
m_{q} {\ddot q_i} = \tilde{\overline{\chi}} - \tilde{\chi}_i = \sum_{j=1}^N \frac{\tilde{\chi}_j - \tilde{\chi}_i + \Phi_j - \Phi_i}{N},
\end{equation}

where $\Phi_i$ and $\Phi_j$ are the external electrostatic potential at the $i$-th and $j$-th atom site, respectively \citep{chen2009charge}. In line with the global charge transfer in EEM, QEq and ES+ (system is approximated as ideal conductor/metal), the polarization due to an external electric field is inherently the polarization of an ideal conductor (metal).

Newton's equation of motion for the nuclei are governed by 
\begin{subequations}
\begin{align}
m_i \boldsymbol{\ddot r_i} &= - \nabla_i U - \sum_{j=1}^N (\lambda +  \frac{\partial U}{\partial q_j}) \nabla_i q_j
\label{eq:qeq_nuclei1}\\
                           &= - \nabla_i U + m_q \sum_{j=1}^N \ddot q_j \nabla_i q_j
\label{eq:qeq_nuclei2}\\
                           &\approx - \nabla_i U
\label{eq:qeq_nuclei3}
\end{align}
\end{subequations}

where in \eqref{eq:qeq_nuclei1} equation \eqref{eq:charge_motion2} is used. Since the charge distribution is equilibrated until the convergence criteria, i.e., set tolerance or precision, is met for each nuclei movement, the second subtrahend is approximately zero ($\ddot q_i \approx 0~\forall i \in [1,N]$). The validity of this approximation is determined by the chosen precision for the charge equilibration.

The original derivation and a discussion of parameters can be found elsewhere \citep{rick1994dynamical}. 

\subsection{Extended Lagrangian method for QTPIE}
\label{ssec:extended_lagrangian_method_for_QTPIE}

While the charge transfer in EEM, QEq, and ES+ is inherently global (ideal conductor, metal), the charge transfer in QTPIE is meant to be local (ideal insulator) \citep{chen2009theory}. This is achieved by the substitution of the neutral electronegativities $\tilde{\chi}_i^0$ with the effective electronegativities
\begin{equation}
\label{eq:QTPIE_effective_electronegativity_definition}
\tilde{\chi}_{\mathrm{eff},i}^0 = \sum_{j=1}^N c_{ij} \frac{\tilde{\chi}_i^0 - \tilde{\chi}_j^0}{N} S_{ij},
\end{equation}

where $c_{ij}$ is a charge independent constant factor and $S_{ij}$ is the $n$s-type overlap integral. In line with QEq, QTPIE makes use of $n$s Slater orbitals. Two suggestions for $c_{ij}$ were made to maintain the correct scale of the atomic electronegativities,
\begin{subequations}
\begin{align}
c_{1,ij} &= \frac{N}{S_{ij}(r_0)},
\label{eq:QTPIE_c1}\\
c_{2,ij} &= c_{2,i} = \frac{N}{\sum_{k=1}^N S_{ik}} .
\label{eq:QTPIE_c2}
\end{align}
\end{subequations}

A derivation is roughly outlined in the introduction and comprehensively described elsewhere \citep{chen2007qtpie, chen2008unified, chen2009theory}.

To interpret QTPIE in the frame of an extended Lagrangian method, we apply equation \eqref{eq:electronegativity_definition2} to equation \eqref{eq:charge_motion2} and substitute the electronegativities
\begin{subequations}
\begin{align}
m_{q} {\ddot q_i} &= - \lambda - \tilde{\chi}_{\mathrm{eff},i}^0 - \sum_{j=1}^N J_{ij}q_j
\label{eq:QTPIE_qLagrangian1}\\
				      &= - \lambda - \sum_{j=1}^N c_{ij} \frac{\tilde{\chi}_i^0 - \tilde{\chi}_j^0}{N} S_{ij} - \sum_{j=1}^N J_{ij}q_j,
\label{eq:QTPIE_qLagrangian2}
\end{align}
\end{subequations}

where the Lagrange multiplier $\lambda$ is determined by enforcing the charge neutrality constraint
\begin{equation}
\label{eq:QTPIE_lambda}
\lambda = -\frac{1}{N}\sum_{i=1}^N \left( \sum_{j=1}^N c_{ij} \frac{\tilde{\chi}_i^0 - \tilde{\chi}_j^0}{N} S_{ij} + \sum_{j=1}^N J_{ij}q_j \right).
\end{equation}
	
The equation resulting by applying equation \eqref{eq:QTPIE_lambda} to equation \eqref{eq:QTPIE_qLagrangian2}, is, however, everything but intuitive to interpret. This can be attributed to the combination of local and global charge transfer, as well as their joint charge neutrality constraint. 

For $c_{ij}=c_{2,ij}$, equation \eqref{eq:QTPIE_lambda} cannot be further simplified due to the asymmetry of $c_{ij}$ ($c_{2,i} \neq c_{2,j}$). Global and local charge transfer interfere with each other while satisfying the charge neutrality constraint.

In case of $c_{ij}=c_{1,ij}$, the complexity of equation \eqref{eq:QTPIE_lambda} can indeed be further reduced due to the symmetry of the overlap integral ($S_{ij}=S_{ji}$ and $c_{ij}=c_{1,ij} = c_{1,ji}$), thus
\begin{equation}
\label{eq:QTPIE_lambda2}
\lambda = -\frac{1}{N}\sum_{i=1}^N\sum_{j=1}^N J_{ij}q_j.
\end{equation}

As a result, local and global charge transfer satisfy the charge neutrality constraint individually, without interfering with each other. 

For an external electric field $\boldsymbol{E} = -\nabla \Phi$, the equation for the fictitious charge motion can be obtained by the appropriate substitution ($\tilde{\chi}_i^0 \rightarrow \tilde{\chi}_i^0 + \Phi_i$ and $\tilde{\chi}_j^0 \rightarrow \tilde{\chi}_j^0 + \Phi_j$).

While QTPIE is meant to approximate all systems as insulators, the combination of local and global charge transfer makes this or any other interpretation less straightforward. It is worthwhile to mention, however, that, similar to QEq, QTPIE may nonetheless be a powerful method.

For the nuclei, the equations of motion are not altered and thus, described by \eqref{eq:qeq_nuclei3} and the surrounding discussion.

Yet, it is not possible to simply constrain the global charge transfer in case of QTPIE. A reduction of the charge transfer range induced by the electrostatic interaction $J_{ij}$, would require a reduced electrostatic interaction $J_{ij}$ itself to maintain the model's consistency. 


\section{Proposal of variable charge models}
\label{sec:proposal_of_variable_charge_models}

In the following, two models are presented which address the issue of global charge transfer inherent in many variable charge models. First, the charge transfer equilibration (QTE) model is introduced. The model is inspired by a comparison of the extended Lagrangian method applied to QEq (EEM, ES+) and QTPIE. Second, an extension to EEM, QEq and ES+ is discussed, which in a methodologically different way enables a better charge transfer for certain surface processes (i.e., adsorption, desorption of single atoms). This method is inspired by the idea of dividing the system into subsystems (e.g., molecules) and defining an individual chemical potential for each \citep{rick1994dynamical}.

\subsection{Charge transfer equilibration (QTE)}
\label{ssec:charge_transfer_equilibration}

In comparison with the previously discussed models, the charge transfer equilibration (QTE) approach differs essentially in two simple, but important decisions.

\subsubsection{Scaled charge transfer variables}
\label{sssec:charge_transfer_variables}

In QTPIE (and SQE) charge transfer variables $q_i$ and $q_{ij}$ are related by

\begin{equation}
\label{eq:QTE_net_charge_transfer_variable}
q_i =  \sum_{j=1}^Nq_{ij},
\end{equation}

while being used in a different ways, respectively \citep{chen2007qtpie, chen2008unified, chen2009theory, mathieu2007split, krykunov2017new}. Here, we go a step further and, in the following, formulate an approach to include the charge transfer $\dot q_{ij}$ as a function of the nuclei distances. 

The transferred charge per unit time $\dot q_{ij}$ corresponds to the rate of the electron transfer $K_\mathrm{ET}$, which (in the classical limit) and without nuclei movement (which are typically anyhow considered to be immobile until the charges are equilibrated) can be described by Marcus theory

\begin{equation}
\label{eq:Marcus_theory}
K_{\mathrm{ET},ij} = \frac{4\pi}{h}\frac{1}{\sqrt{4\pi k_\mathrm{B}T\lambda}} \boldsymbol |H_{ij}|^2 \exp{-\frac{(\lambda+\Delta G)^2}{4k_\mathrm{B}T}},
\end{equation}

where $h$ is Plank's constant, $k_\mathrm{B}$ is Boltzmann's constant, $T$ is the absolute temperature, $\lambda$ is the reorganization energy, $H_{ij}$ is the electronic coupling between the initital and final state, $\Delta G$ is the respective change of the Gibbs free energy \citep{marcus1956electrostatic, marcus1956theory,liu2005ionization}. The electronic coupling can be roughly approximated by the respective overlap integral $S_{ij}$ \citep{schuster2004long}. Thus, $K_{\mathrm{ET},ij}$ is approximately proportional to $S_{ij}^2$. We propose therefore the following ansatz for the transferred charge per unit time

\begin{equation}
\label{eq:QTE_ansatz}
\dot q_{ij} = c_{ij} S_{ij}^m\dot p_{ij},
\end{equation}

where $c_{ij}$ is a constant for the fictitious charge motion, $m$ is the exponent of the overlap integral $S_{ij}$, $p_{ij}$ may be interpreted as a constrained charge transfer variable. 

To obtain a model which may be applicable for various RMD potentials, we do not specify the orbital type of the overlap integrals on purpose. We believe, that the orbital (e.g., $n$s Slater type orbital or a linear combination of Gaussian type orbitals) should be chosen consistent with the particular RMD potential of interest. This will be more thoroughly discussed in Section~\ref{sec:discussion}.

As pointed in the beginning of Section~\ref{ssec:extended_lagrangian_method_for_EEM_QEq_and_ES+}, due to the different time scales of the nuclei and the electrons, the respective equations of motion are usually solved sequentially and not in parallel. The charges are equilibrated for each nuclei displacement. Furthermore, the geometry of the nuclei can be thought of as static (frozen) background for the fictitious charge motion. As a consequence, the overlap integral $S_{ij}$ is a constant for the fictitious charge motion, too.

The transformation from atomic to charge transfer variables can thus be described by combining equation \eqref{eq:QTE_net_charge_transfer_variable} and \eqref{eq:QTE_ansatz} to

\begin{subequations}
\begin{align}
&\dot q_i       = \sum_{j=1}^N c_{ij}S_{ij}^m \dot p_{ij}.
\label{eq:QTE_charge_transfer_velocity}
\end{align}
\end{subequations}

The differentiation provides 

\begin{subequations}
\begin{align}
&\ddot q_i      = \sum_{j=1}^N c_{ij}S_{ij}^m \ddot p_{ij}
\label{eq:QTE_charge_transfer_acceleration}
\end{align}
\end{subequations}

and the integration results in

\begin{subequations}
\begin{align}
&q_i - q_i(t=0) = \sum_{j=1}^N c_{ij}S_{ij}^m p_{ij} - \int \sum_{j=1}^N p_{ij} \frac{\mathrm{d}(c_{ij}S_{ij}^m)}{\mathrm{d}t}\mathrm{d}t
\label{eq:QTE_charge_transfer_variable0}\\
\Leftrightarrow &q_i - q_i(t=0)	=  \sum_{j=1}^N c_{ij}S_{ij}^m p_{ij} - \sum_{j=1}^N p_{ij} \int \mathrm{d}(c_{ij}S_{ij}^m).
\label{eq:QTE_charge_transfer_variable}
\end{align}
\end{subequations}

In equation \eqref{eq:QTE_charge_transfer_variable0} one has to consider, that $c_{ij}S_{ij}^m$ evolves on the nuclei time scale, for which individual displacement $p_{ij}$ is assumed to be constant. The subtrahend can be understood as charge transport due to the nuclei rearrangement, which eventually either weakens or strengthens the individual charge transfer between atom pairs. Thus, it represents the history of all previous charge exchanges. Hence, the net charge of separated subsystems is a function of its current a well as preceding atom configurations. Charge neutrality for each subsystem can be enforced by extending QTE to QTE$^+$: It only requires to set to zero the particular term referenced, respectively the subtrahend of equation \eqref{eq:QTE_charge_transfer_variable0} or \eqref{eq:QTE_charge_transfer_variable}. This procedure is useful when the employed interaction potential struggles to accurately describe the ionization state of the respective subsystems.

\subsubsection{Extended Lagrangian method for QTE and QTE$^+$}
\label{sssec:extended_lagrangian_method_for_QTE}

The second major difference in comparison to EEM, QEq, ES+ and QTPIE relates to the considered quantity. Specifically, we evolve the (constrained) interatomic charge transfer instead of the atomic charge distribution in time. While most other models enforce charge neutrality, $\sum_{i=1}^N q_i = 0$, here it is charge conservation that is considered, $\sum_{i=1}^N \dot q_i = 0$. The initial net charge of the system will therefore be kept constant during the simulation. Since the method of the Lagrange multiplier leads to a global charge transfer, $c_{ij}$ is required to satisfy the charge conservation constraint. The corresponding equation of motion is evaluated through

\begin{subequations}
\begin{align}
m_{q} \ddot p_{ij} &= - \frac{\partial U}{\partial p_{ij}} - \sum_{l=1}^N \nabla_l U \frac{\partial \boldsymbol r_l}{\partial p_{ij}}
\label{eq:QTE_Lagrangian1}\\
                    &= - \frac{\partial U}{\partial p_{ij}} 
\label{eq:QTE_Lagrangian1sub}\\
					 &= - \frac{\partial U}{\partial q_i} \frac{\partial q_i}{\partial p_{ij}} - \frac{\partial U}{\partial q_j} \frac{\partial q_j}{\partial p_{ij}}
\label{eq:QTE_Lagrangian3}\\			
					 &= - \tilde{\chi_i} \frac{\partial q_i}{\partial p_{ij}} + \tilde{\chi_j} \frac{\partial q_j}{\partial p_{ji}} 			 
\label{eq:QTE_Lagrangian5}\\
					 &= - \tilde{\chi_i} \sum_{k=1}^N c_{ik}S_{ik}^m \frac{\partial p_{ik}}{\partial p_{ij}} + \tilde{\chi_j} \sum_{k=1}^N c_{jk}S_{jk}^m \frac{\partial p_{jk}}{\partial p_{ji}} 			 
\label{eq:QTE_Lagrangian6}\\
					 &= - \tilde{\chi_i} \sum_{k=1}^N c_{ik}S_{ik}^m \delta_{jk} + \tilde{\chi_j} \sum_{k=1}^N c_{jk}S_{jk}^m \delta_{ik}			 
\label{eq:QTE_Lagrangian7}\\
					 &= - c_{ij}S_{ij}^m \tilde{\chi_i} + c_{ji}S_{ji}^m \tilde{\chi_j}
\label{eq:QTE_Lagrangian8}\\
					 &=  c_{ij}S_{ij}^m \left( \tilde{\chi_j} - \tilde{\chi_i} \right).
\label{eq:QTE_Lagrangian9}
\end{align}
\end{subequations}

Since the nuclei sites are constant during the fictitious charge motion, the second subtrahend in equation \eqref{eq:QTE_Lagrangian1} vanishes. For equation \eqref{eq:QTE_Lagrangian5} we make use of equation \eqref{eq:electronegativity_definition1} using $\tilde{\chi}_k = \frac{\partial U}{\partial q_k}$ and the symmetry of the charge transfer variables $q_{ij}=-q_{ji} \Leftrightarrow p_{ij}=-p_{ji}$; for equation \eqref{eq:QTE_Lagrangian6} we apply equation \eqref{eq:QTE_charge_transfer_variable}. For equation \eqref{eq:QTE_Lagrangian9} we utilize the symmetry of the overlap integral $S_{ij}=S_{ji}$ and claim $c_{ij}=c_{ji}$. This means that the fictitious charge force $f_{q,ij}$, which is defined by the right hand side of equation (\ref{eq:QTE_Lagrangian9}), satisfies Newton's third law ($f_{q,ij} = -f_{q,ji}$). Due to this symmetry, the conservation of charge is independent from $c_{ij}$ and consequently always fulfilled. This can be tested by the summation over $i,j \in [1,N]$. 

The next goal is to transform the charge transfer variables back to atomic charge variables, starting from equation \eqref{eq:QTE_Lagrangian9}. First, we use equation \eqref{eq:QTE_ansatz} to obtain the charge transfer variables

\begin{equation}
\label{eq:QTE_Lagrangian_transformed1}
m_q \ddot q_{ij} = c_{ij}^2 S_{ij}^{2m} \left( \tilde{\chi_j} - \tilde{\chi_i} \right).
\end{equation}

As outlined in the beginning of Section~\ref{sssec:charge_transfer_variables}, in this approach, we make use of the approximated proportionally of the charge transfer with regard to the square of the overlap integral. Thus, we choose $m=1$. 

To proceed, the summation over $j \in [1,N]$ is performed

\begin{equation}
\label{eq:QTE_Lagrangian_transformed2}
\sum_{j=1}^N m_q \ddot q_{ij} = \sum_{j=1}^N  c_{ij}^2 S_{ij}^{2} \left( \tilde{\chi_j} - \tilde{\chi_i} \right).
\end{equation}

We then make us of equation \eqref{eq:QTE_charge_transfer_acceleration}, yielding
\begin{equation}
\label{eq:QTE_Lagrangian_transformed3}
m_{q} \ddot q_i = \sum_{j=1}^N  c_{ij}^2 S_{ij}^{2} \left( \tilde{\chi_j} - \tilde{\chi_i} \right).
\end{equation}

The last unknown, $c_{ij}$, is required to fulfill three specifications: i) symmetry $c_{ij}=c_{ji}$, ii) agreement with QEq in the limiting case of global charge transfer ($S_{ij}=1$), iii) sustaining the local charge transfer. While $c_{ij}$ can be defined in multiple ways, we propose here the following heuristic form
\begin{equation}
\label{eq:QTE_cij_definition}
c_{ij}^2 = \frac{2}{ \sum_{k=1}^N S_{ik}^2 + S_{jk}^2 }.
\end{equation}

The final equations of motion for atomic charge are then as follows
\begin{equation}
\label{eq:QTE_charge_motion}
m_{q} \ddot q_i = \sum_{j=1}^N  \frac{2 S_{ij}^2 \left( \tilde{\chi_j} - \tilde{\chi_i} \right)}{ \sum_{k=1}^N S_{ik}^2 + S_{jk}^2 }.
\end{equation}

The electronegativity difference can be interpreted as upper limit for the instantaneous fictitious charge force. For the latter the quotient ${2S_{ij}^2}/{\sum_{k=1}^N S_{ik}^2 + S_{jk}^2}$ functions as a weight. The overlap integral is put in relation to the sum of all neighboring overlap integrals, yielding the respective fraction. In total, the interatomic charge transfer is constrained by the orbital overlap distribution, resembling a network for the fictitious current.

With an external electric field $\boldsymbol{E} = -\nabla \Phi$, the charge equations of motion become
\begin{equation}
\label{eq:QTE_extternal_field}
m_{q} \ddot q_i = \sum_{j=1}^N  \frac{2 S_{ij}^2 \left( \tilde{\chi_j} - \tilde{\chi_i} + \Phi_j - \Phi_i \right)}{ \sum_{k=1}^N S_{ik}^2 + S_{jk}^2},
\end{equation}

where $\Phi_i$ and $\Phi_j$ are the electrostatic potential at the $i$-th and $j$-th atom site, respectively. 

For the $i$-th nuclei, the equations of motion are

\begin{subequations}
\begin{align}
m_i \boldsymbol{\ddot r_i}   &= - \nabla_i U - \sum_{j=1}^N\sum_{k=1}^N \frac{\partial U}{\partial p_{jk}} \nabla_i p_{jk} 
\label{eq:QTE_nuclei_motion_1}\\
                             &= - \nabla_i U - m_q \sum_{j=1}^N\sum_{k=1}^N \ddot p_{jk} \nabla_i p_{jk} 
\label{eq:QTE_nuclei_motion_2}\\
                             &\approx - \nabla_i U,
\label{eq:QTE_nuclei_motion_3}
\end{align}
\end{subequations}

where equation \eqref{eq:QTE_Lagrangian1sub} is used in equation \eqref{eq:QTE_nuclei_motion_1}. Since the charges are equilibrated ($\ddot p_{jk} \approx 0~\forall j,k \in [1,N]$) for each nuclei movement, the second subtrahend in equation \eqref{eq:QTE_nuclei_motion_2} vanishes.

A thorough discussion on parameters and how to set up the extended Lagrangian method in general (e.g., convergence criteria) can be found elsewhere \citep{rick1994dynamical}. 

To generalize this model without applying the extended Lagrangian method (e.g., in the frame of ReaxFF), the $i$-th equation of the system of $N$ linear equations can be obtained by setting $\ddot q_i$ to zero in equation \eqref{eq:QTE_extternal_field},
\begin{equation}
\label{eq:QTE_set_of_lin_equations}
\sum_{j=1}^N  \frac{2 S_{ij}^2 \left( \tilde{\chi_j} - \tilde{\chi_i} + \Phi_j - \Phi_i \right)}{\sum_{k=1}^N S_{ik}^2 + S_{jk}^2} = 0.
\end{equation}

One may use $\tilde{\chi_i} = \tilde{\chi_i}^0 + \sum_{j=1}^N q_j J_{ij}$ (i.e., equation \eqref{eq:electronegativity_definition2}) to determine the electronegativity.

If no external electric field is applied, $\Phi_i$ and $\Phi_j$ have to be set to zero.

\subsection{Extension to EEM, QEq and ES+ for specific surface processes}
\label{ssec:extension_to_QEq}

QTE and QTE$^+$ are proposed to inherently account for varying charge transfer conditions encountered in numerous surfaces processes (e.g., dissociation, recombination, desorption, adsorption, sputtering, deposition, fragmentation). In the following, in contrast, an alternative to the preceding model is described, which enables EEM, QEq and ES+ to treat a few surface processes (i.e., adsorption, desorption of single atoms) appropriately, too. In the introduction an approach was mentioned, where the system is divided into subsystems (e.g., molecules) and multiple chemical potentials (one per subsystem) are defined \citep{rick1994dynamical}. Here, we simply divide our system into $1+N$ subsystems. The first one is the surface with the chemical potential $\mu$. The other $n$ ones are determined by setting each atom's charge individually to zero, defining $\mu_{0,i}$ chemical potentials. An atom far away from the surface slab is therefore enforced to be neutral. At last, a tapering function $f_i(r_i$) (e.g., the Tersoff cutoff function used in COMB3 or the tapering function used in ReaxFF \citep{liang2013classical, van2001reaxff}) is utilized to create a smooth transition from the first to the second chemical potential. The transition region may be defined by the lower and upper bound $h_{\mathrm{lo}}$ and $h_{\mathrm{hi}}$, respectively,  along the surface normal $\boldsymbol n$. The tapering function must satisfy $f_i(r_i) = 1$ for $r_i \leq h_\mathrm{lo}$ and $f_i(r_i) = 0$ for $r_i \geq h_\mathrm{hi}$. We suggest to choose $h_\mathrm{lo}$ slightly above the last atom in $\boldsymbol n$ direction. If the potential of interest (e.g., ReaxFF, COMB3) employs a cutoff radius $r_\mathrm{c}$, $h_\mathrm{hi}$ can be simply chosen to be $h_\mathrm{hi} = r_\mathrm{c}+h_\mathrm{lo}$. For deposition or sputtering simulation, however, it may be good practice to regularly adjust both bounds, due to the variation of the surface height in $\boldsymbol n$ direction. 

Using equation \eqref{eq:chemical_potential_final}, the $i$-th linear equation for the atomic charge distribution becomes
\begin{subequations}
\begin{align}
&f_i \mu + (1-f_i) \mu_{0,i} = 0
\label{eq:extension_set_of_lin_equations1}\\
\Leftrightarrow &f_i \left( \sum_{j=1}^N  \frac{ \tilde{\chi}_j - \tilde{\chi}_i}{N} \right) + (1 - f_i) \left( \sum_{j=1}^N  \frac{  \tilde{\chi}_j|_{q_i=0} - \tilde{\chi}_i|_{q_i=0}  }{N} \right) = 0.
\label{eq:extension_set_of_lin_equations2}
\end{align}
\end{subequations}


\section{Mirror boundary condition}
\label{sec:mirror_boundary_condition}

Surface simulations typically employ a slab, which often is created by cleaving the respective bulk system perpendicular to the surface normal $\boldsymbol n$. This slab is often meant to be an approximation for the surface and bulk system of interest, taking the second (lower) surface as necessary circumstance. Thus, the interaction between both surfaces is in many cases highly undesirable. This issue is typically addressed by creating a thicker slab to suppress the respective interaction. The increased thickness, however, consequently enlarges the computational cost. Furthermore, the atoms which belong to the lower surface are usually not evolved in time (frozen), avoiding unnecessary computations and fixing the slab position. In case of RMD simulations, specifically variable charge models, the lower surface imposes nonetheless additional computational effort. While the nuclei at the lower surface can be kept frozen, the charge, however, has to be equilibrated throughout the total system. This circumstance is substantial, if one considers that charge equilibration is usually one of the most time consuming aspects of RMD simulations.

In the following, a mirror boundary condition (MBC) is described which overcomes this computational burden. A symmetry for the lower bound of the simulation box in $\boldsymbol n$ direction is introduced, which enforces all atoms beneath the specified height $h_\mathrm{hi}$ to interact with each other as well as with their mirror images. Mirror images are replicates of the original atoms with inverted $\boldsymbol r \cdot \boldsymbol n$ coordinate. The respective domain will be referred to as mirroring zone. For RMD potentials that make use of a cutoff radius $r_\mathrm{c}$ (e.g., ReaxFF, COMB3), $h_\mathrm{hi}$ should be chosen to be slightly larger than $r_\mathrm{c}$. It is recommended to exclude the nuclei below $h_\mathrm{hi}$ from the time integration (frozen) to avoid introducing artifacts into the dynamics. Overall, the MBC corresponds to a homogeneous Neumann boundary condition for the total potential energy $U$,
\begin{equation}
\label{eq:MBC_Neumann}
\frac{\partial U}{\partial n} = 0.
\end{equation}

\begin{figure}[t]
\begin{center}
\resizebox{8cm}{!}{
\includegraphics[width=8cm]{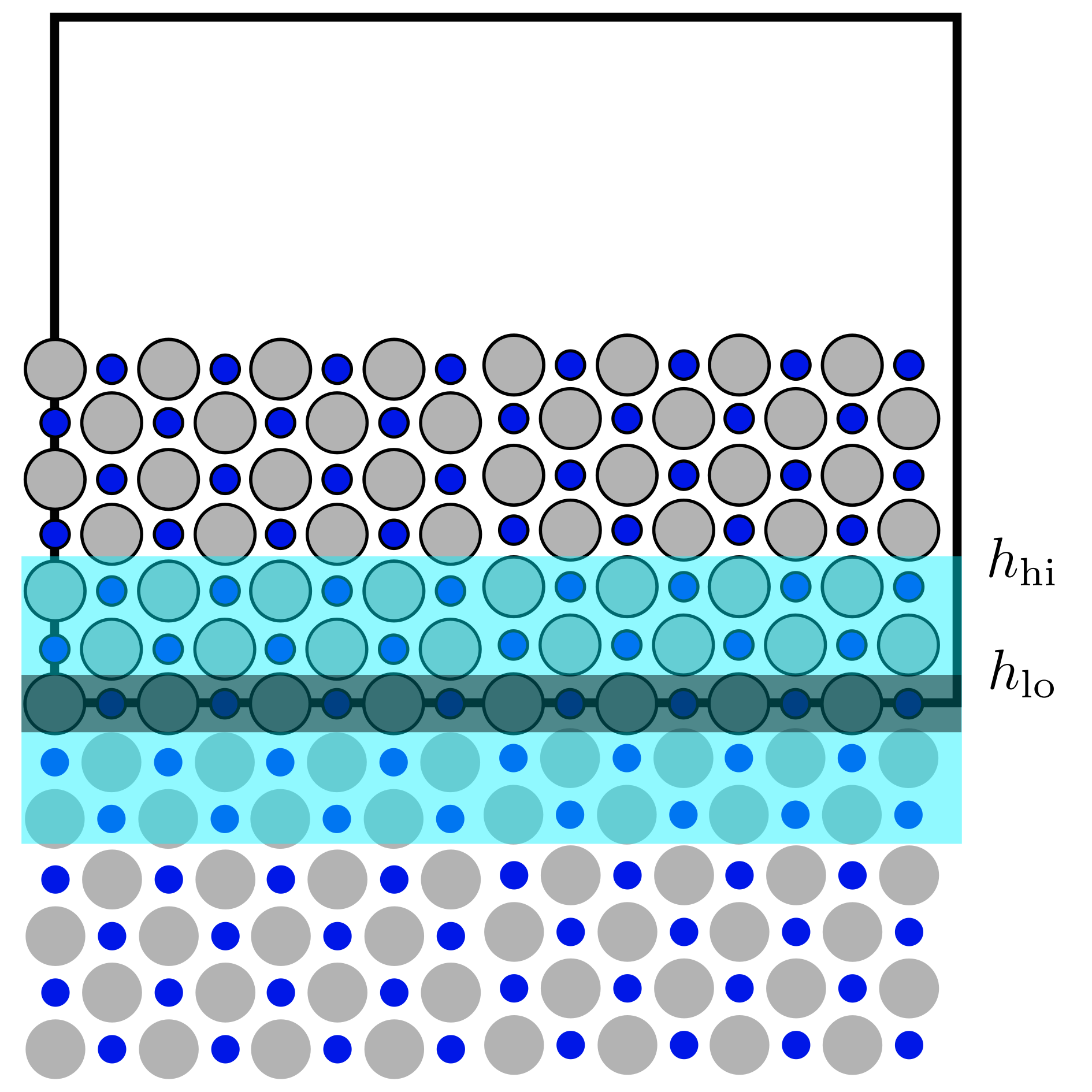}
}
\caption{Schematic representation of the mirror boundary condition for B1 TiN (rocksalt lattice structure). Titanium and nitrogen atoms are coloured grey and blue, respectively. The depicted slab width, thickness, $h_\mathrm{hi}$ and $h_\mathrm{lo}$ are meant to illustrate the concept and do not correspond to appropriate choices for a RMD simulation. From top to bottom: atoms, frozen atoms, frozen atoms without mirror images, mirror images of frozen atoms, mirror images of atoms.}
\label{fig:mbc}
\end{center}
\end{figure}

There are many crystal structures, however, that do not allow for this kind of introduced symmetry. To enlarge the number of possible materials, a lower bound $h_\mathrm{lo}$ for the mirroring zone is defined. Care has to be taken, to include only the first layer of atoms, which should be positioned very close the lower bound of the simulation box in $\boldsymbol n$ direction. Then, the atoms in this first layer do not have any mirror images (while interacting with the other atoms and their mirror images). Though, this first layer of atoms has to be charge neutral to sustain charge conservation (or charge neutrality) of the total system. Unfortunately, this reduces the number of potential crystals structures to be studied. The MBC is nonetheless beneficial for the remaining ones (e.g., rocksalt lattice structure). A schematic of the MBC for B1 TiN (rocksalt lattice structure) is presented in Figure~\ref{fig:mbc}.

Whenever a global quantity (e.g., net charge, total energy, stress tensor) is computed as a function of all atom properties, care has to be taken to account for the mirror images appropriately. For example, when checking for charge neutrality (or charge conservation), the net charge $Q$ of the system is described by

\begin{equation}
\label{eq:MBC_netQ}
Q = \sum_{i=1}^N q_i  \left(1 + \left(\boldsymbol r_i \cdot \boldsymbol n > h_\mathrm{lo}\right)\right)
\end{equation}


\section{Implementation}
\label{sec:implementation}

The implementation of the preceding models is often straightforward. In the following, however, we are going to highlight a few advisable aspects. All models are implemented in the widely used open-source code LAMMPS \citep{plimpton1995fast}. 

First, a few basics concerning the parallelization in LAMMPS are provided. Each processor handles a subdomain of the total simulation box. Ghost atoms are used to allow for atoms to interact with each other while being owned by different processors. This is in particular useful for boundary conditions. Each atom keeps all its ghost atoms up to date (e.g., concerning atom site and charge). The ghost atoms, on the other hand, transfer information (e.g., about the experienced force and fictitious charge force) back to the real atoms. The migration from one to another processor can be triggered if an atom leaves the respective sub-domain. The migration, however, will only be executed, when the neighbor list is built the next time.

The implementation of the extended Lagrangian method in combination with a Verlet algorithm may cause harm, when an atom migrates from one to another processor. The latter may have no information about the fictitious charge force of the atom, which consequently leads to a violation of the charge conservation (and charge neutrality). Yet, this issue can straightforwardly be dealt with: When the neighbor list is rebuilt during the nuclei timestep, the subsequent fictitious charge motion are restarted ($\ddot q_i = \dot q_i = 0~\forall~i\in[1,N]$) from the current atomic charge distribution $q_i$. Even when its unnecessary (no atom migration between processors), this procedure does no harm to the respective fictitious charge dynamics. If EE was exactly fulfilled, $\ddot q_i$ and $\dot q_i$ would equal 0 anyhow for all nuclei motions. However, EE is only met up to a set tolerance (precision), so that some residual fictitious charge acceleration as well as velocity remain during the nuclei time step. This initial guess for the following charge equilibration only means that less iterations are required to meet convergence. However, neighbor lists are rarely rebuilt, so that this does not lead to significant overhead.

For the application of QTE (and QTPIE), the overlap integrals have to be evaluated only once for each charge equilibration run (nuclei are frozen).

Though, the efficient implementation of QTE$^+$ is more challenging since certain necessities are at least to some extent in conflict with the way LAMMPS operates. One has to keep track of the scaled transferred charge $p_{ij}$ between all atoms as well as the change in the factor $c_{ij}S_{ij}$ (see equation \eqref{eq:QTE_charge_transfer_variable}). So all processors may eventually demand access to this information for any pair of atoms. But typically, in LAMMPS, each processor stores only the information for its currently owned atoms and may access the information provided by ghost atoms, which are provided by neighbored processors. This issue is solved by having the local data gathered among and then stored by all processors, indexed with the atom's global tag (which therefore has to remain constant, not compressed throughout the simulation). The required communication between all processors is computationally expensive and should therefore only be executed when absolutely necessary (i.e., when atoms migrate from one processor to another during a neighbor list rebuilt). Subsequently, the local information are reset and updated correspondingly. Concerning memory, this solution is rather inefficient. However, since RMD studies are usually computationally demanding, often smaller system sizes are chosen. Consequently, an even partially inefficient memory management shouldn't become a problem. The procedure is as follows. First, if the neighbor list is rebuilt, the local information (i.e., $c_{ij}S_{ij}$ and $p_{ij}$) are gathered and distributed among all processors. Each of which resets its data from the preceding timestep and updates the data for its currently owned atoms. Second, the atomic charges are adjusted to account for the change in the overlap integral by the neglected subtrahend in equation \eqref{eq:QTE_charge_transfer_variable}. Third, the charges are equilibrated while tracking the scaled transferred charge $p_{ij}$ locally at the same time.

The implementation of the MBC can be easily achieved utilizing the ghost atom concept in LAMMPS. When setting up the simulation, each atom with $\boldsymbol r \cdot \boldsymbol n \in [h_\mathrm{lo},~h_\mathrm{hi}]$ is used to create its own mirror image as ghost atom (the respective processors are marked with a mirror flag). Naturally, the atoms will then update their mirror images. Subsequently, the processors share the ghost atoms with each other. Thus, the interactions between atoms and mirror images will be performed self-consistently.

When Newton's 3rd law is used to reduce the number of computations, the implementation of the MBC becomes more complicated. In LAMMPS, the application of Newton's 3rd law leads to an iteration over half the neighbor list, using a global tag per atom (processor independent). The standard procedure deals efficiently with this task while at the same time, handling the self interaction of atoms in case of very small periodic systems appropriately. 

For the MBC, if $h_\mathrm{lo}$ differs from the lower bound of the simulation domain in the surface normal direction $\boldsymbol n$, three exceptions have to be considered additionally. As mentioned in Section~\ref{sec:mirror_boundary_condition}, nuclei are meant to be excluded from the time integration (immobile). Thus, only the scalar fictitious charge forces $f_{q,ij}$ have to be adapted. They are defined by 
\begin{equation}
\label{eq:implementation_charge_force}
m_q \ddot q_i = \sum_{j}^N f_{q,ij}
\end{equation}

and can be determined in comparison with equations \eqref{eq:charge_motion_final}, \eqref{eq:QTPIE_qLagrangian2} and \eqref{eq:QTE_charge_motion} for EEM (QEq, ES+), QTPIE and QTE, respectively. The fictitious charge forces at the MBC have to be treated as follows
\begin{subequations}
\begin{align}
&f_{q,ji} \rightarrow 2f_{q,ji}& ~&\mathrm{if}&~ &(h_\mathrm{lo} < \boldsymbol r_i \cdot \boldsymbol n < h_\mathrm{hi}) ~\mathrm{and}~ (|\boldsymbol r_j \cdot \boldsymbol n| < h_\mathrm{lo})
\label{eq:implementation_adapted_charge_force2}\\
&f_{q,ji} \rightarrow 0& ~&\mathrm{if}&~ &(|\boldsymbol r_i \cdot \boldsymbol n| < h_\mathrm{lo}) ~\mathrm{and}~ (h_\mathrm{lo} < -\boldsymbol r_j \cdot \boldsymbol n < h_\mathrm{hi}),
\label{eq:implementation_adapted_charge_force3}
\end{align}
\end{subequations}

As mentioned in Section~\ref{sec:mirror_boundary_condition}, the net charge of the atoms which satisfy  $(|\boldsymbol r_i \cdot \boldsymbol n| < h_\mathrm{lo})$ has to equal zero to avoid a violation of the charge conservation (and charge neutrality). Furthermore, the implementation of the MBC should be verified by at least checking for charge neutrality (or charge conservation) as described in Section~\ref{sec:mirror_boundary_condition}.


\section{Validation}
\label{sec:validation}

This section is devoted to the demonstration of the validity and limitations of the fluctuating charge models QEq, QTPIE, exQEq, QTE and QTE$^+$. Thus, it is necessary to specify the interatomic interactions (i.e., the electrostatic interactions) to compute the equilibrated atomic charge distribution. Here, the variable charge models are used in combination with the RMD potential COMB3. Within the particular formalism, the electrostatic interactions are based on a point charge distribution of nuclei and and a 1s Slater-type electron charge distribution. A thorough discussion of the model and its parameters can be found elsewhere \citep{liang2013classical,liang2012variable}. To couple QTE or QTPIE with COMB3, the exponent $\zeta$ of the 1s Slater-type orbitals (STOs), which describe the radial decay in space, is allowed to differ from the exponent $\zeta^\mathrm{ov}$, which is solely used for the computation of the overlap integrals $S_{ij}$. The latter is obtained by fitting the single element 1s1s overlap integral to the respective $n$s$n$s overlap integral, where $n$ is the principle quantum number of the valence electron. The 1s1s overlap integral can be expressed by two closed formula: for $\zeta^\mathrm{ov}_1=\zeta^\mathrm{ov}_2$,

\begin{equation}
\label{eq:overlap1s1sSameZeta}
S_{ij}^\mathrm{1s1s} = \left(1.0+\rho+\frac{1}{3} \rho^2 \right)e^{-\rho}
\end{equation}

and for $\zeta_1^\mathrm{ov}\neq\zeta_2^\mathrm{ov}$,

\begin{equation}
\label{eq:overlap1s1sZeta}
S_{ij}^\mathrm{1s1s} =  \frac{\sqrt{1-\tau^2}}{\tau \rho} \left(-(1-\kappa) (2 (1+\kappa)+\rho_i) e^{-\rho_i}+(1+\kappa)(2(1-\kappa)+\rho_j) e^{-\rho_j} \right),
\end{equation}

with $\rho = \frac{\zeta_i+\zeta_j}{2}r_{ij},\tau = \frac{\zeta_i-\zeta_j}{\zeta_i+\zeta_j}, \kappa = \frac{1}{2} (\tau+\frac{1}{\tau}), \rho_i = \zeta_i r_{ij}$ and $\rho_j = \zeta_j r_{ij}$ \citep{roothaan1951study,mulliken1949formulas}. To model the orbital overlap in consistence with the vanishing atom interaction beyond the cutoff radius in COMB3, the Tersoff-cutoff function $f_C(r_{ij})$ is used to taper the overlap integrals,

\begin{equation}
\label{eq:overlapTapered}
S_{ij}^\mathrm{1s1s} \rightarrow f_C(r_{ij}) S_{ij}^\mathrm{1s1s}.
\end{equation}

The lower and upper bound are chosen to be 9 \r A and 11 \r A, respectively. 

The computation of the $n$s$n$s overlap integrals, require numerical integration and is described elsewhere \citep{rosen1931calculation,chen2009theory}. As a result, the fitted parameters for most of the elements, which are currently available in the frame of COMB3, are listed in table \ref{eq:tabZetaQTECOMB3}. Missing or new parameters can either be obtained by following the beforehand mentioned procedure or using $\zeta^\mathrm{ov}=\zeta$ as a default value. 

\begin{table}
\begin{tabular}{lc}
\hline
element & $\zeta^\mathrm{ov}$ (\r A$^{-1}$)  \\  
\hline
C & 1.036 \\
N & 1.239 \\
O & 1.447 \\
Al & 0.668 \\
Ti & 0.469 \\
Ni & 0.564 \\
Cu & 0.578 \\
Zn & 0.591 \\
\hline
\end{tabular}
\caption{The 1s STO exponent $\zeta^\mathrm{ov}$ which is fitted to reproduce the $n$s$n$s single elemental overlap integral for various elements, which are used in the frame of COMB3.}
\label{eq:tabZetaQTECOMB3}
\end{table}

In the following, the fluctuating charge models QEq, QTPIE, exQEq, QTE and QTE$^+$ are compared to each other for a variety of molecular statics (0 K) simulations of surface processes as well as externally applied electric fields. 

\subsection{Surface processes}
\label{ssec:Surface processes}

\begin{figure}[t]
\begin{minipage}[t]{0.49\linewidth}
\resizebox{8cm}{!}{
\includegraphics[width=8cm]{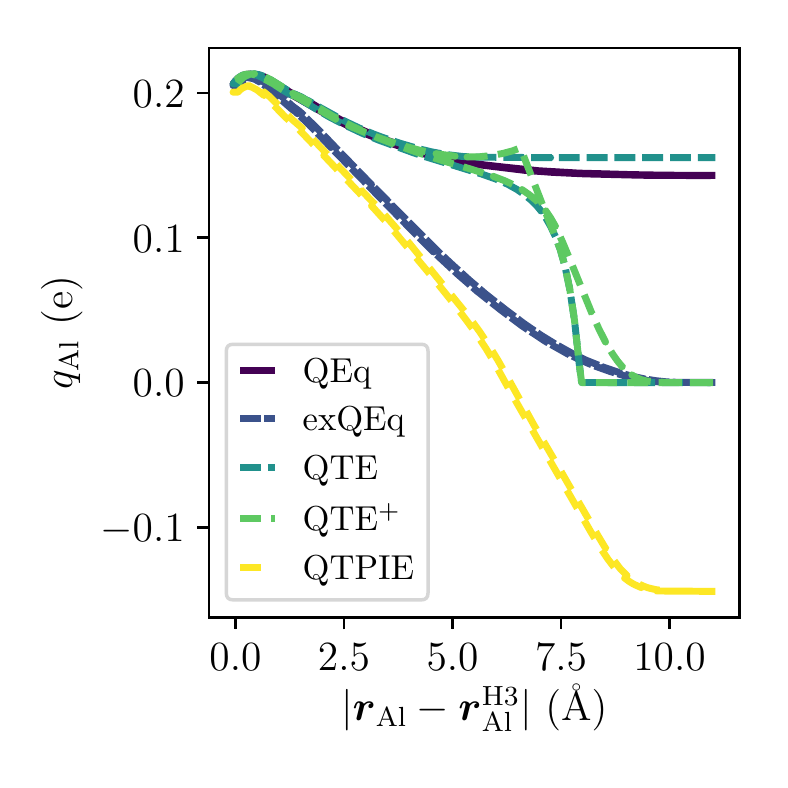}
}
\caption{The charge of an aluminium atom $q_\mathrm{Al}$ is displayed against the distance to the hollow adsorption site $\boldsymbol{r}_\mathrm{Al}^\mathrm{H3}$ on the AlN($00\overline{1}$) surface.}
\label{fig:qAlNAl}
\end{minipage}
\hfill
\begin{minipage}[t]{0.49\linewidth}
\resizebox{8cm}{!}{
\includegraphics[width=8cm]{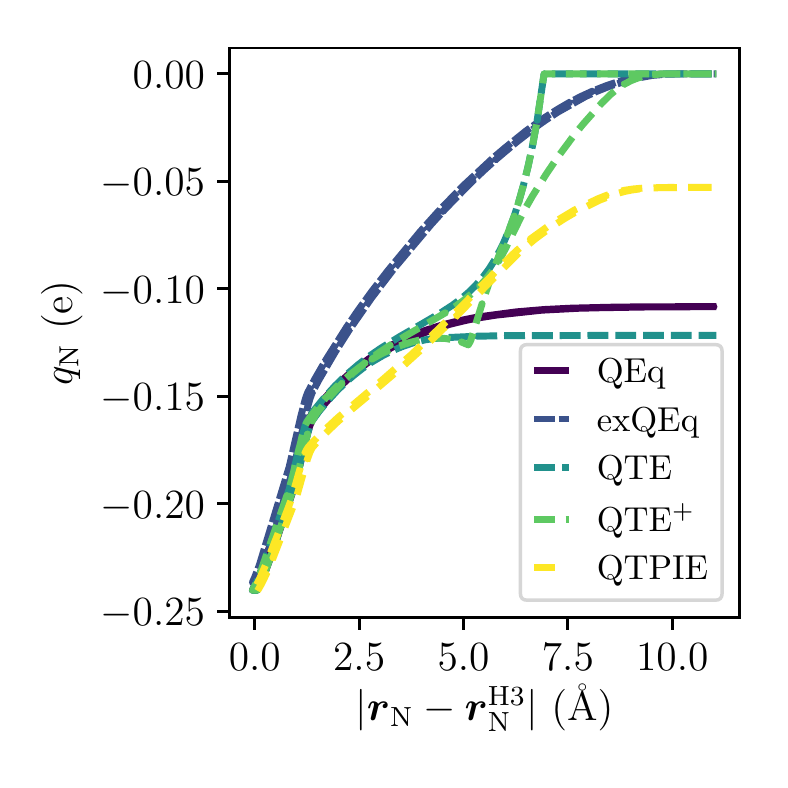}
}
\caption{The charge of a nitrogen atom $q_\mathrm{N}$ is displayed against the distance to the hollow adsorption site $\boldsymbol{r}_\mathrm{N}^\mathrm{H3}$ on the AlN($001$) surface.}
\label{fig:qAlNN}
\end{minipage}
\end{figure}

First, the charge variation during the adsorption and desorption process of an Al atom at the AlN(00$\overline{1}$) hollow surface site and a N atom at the hollow  AlN(001) surface site are presented in Figure~\ref{fig:qAlNAl} and \ref{fig:qAlNN}, respectively. For the adatom at the adsorption site, we consider QEq as reference for the other models (no separated subsystems). Though, for larger distances QEq predicts a charged atom for either adsorption or desorption. As mentioned previously, the underlying global charge transfer is artificial and nonphysical. The exQEq model is designed to deal with this issue and and therefore yields neutral atoms for larger distances. The QTE model starts off with a neutral atom far from the surface. While lowering to the surface, the overlap integral and consequently the charge transfer is increased until QTE agrees exactly with QEq at the surface. However, during the desorption process, the adatom remains as a charged particle and does not become neutral. QTE$^+$ is meant to resolve with this potential issue by negating the charge transfer from previous timesteps. As a result, the desorbed atom becomes neutral when leaving the surface. QTPIE fails to predict a neutral charge state while being applied to two separated subsystems. While decreasing the distance to the hollow adsorption site, the model agrees more and more with QEq. However, unlike for QTE or QTE$^+$, there is no exact agreement. 

\begin{figure}[t]
\begin{minipage}[t]{0.49\linewidth}
\resizebox{8cm}{!}{
\includegraphics[width=8cm]{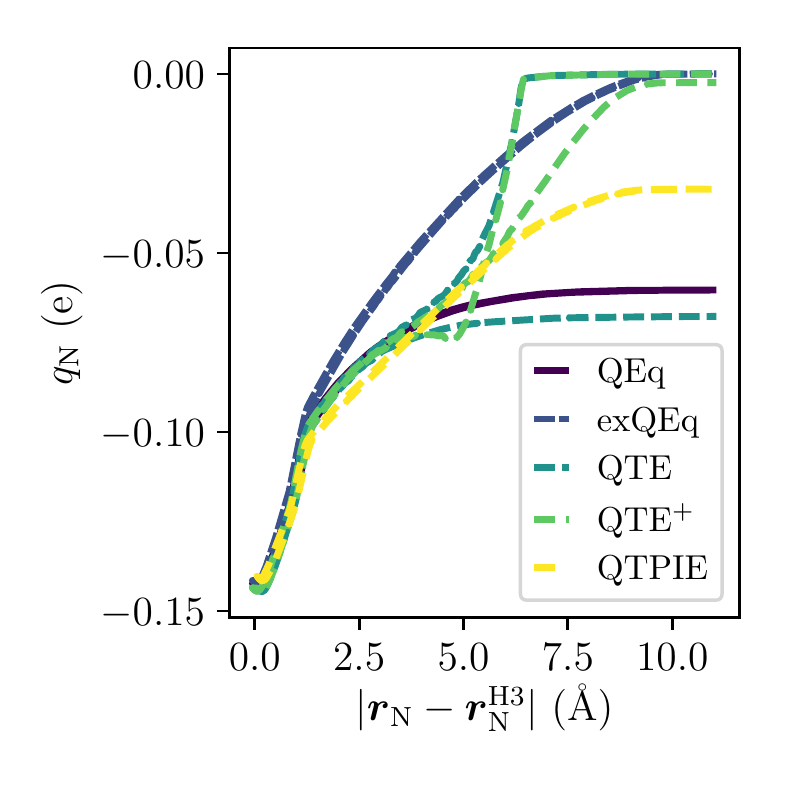}
}
\caption{The charge of the lower nitrogen atom $q_\mathrm{N}$, which is bonded to another, upper nitrogen atom (forming a nitrogen molecule), is displayed against the distance to the hollow adsorption site $\boldsymbol{r}_\mathrm{N}^\mathrm{H3}$ on the AlN($001$) surface. The $N_2$ bond ($|\Delta \boldsymbol r_\mathrm{N_2}|$ = 1.1 \r A) is parallel to the surface normal.}
\label{fig:qAlNN2_1}
\end{minipage}
\hfill
\begin{minipage}[t]{0.49\linewidth}
\resizebox{8cm}{!}{
\includegraphics[width=8cm]{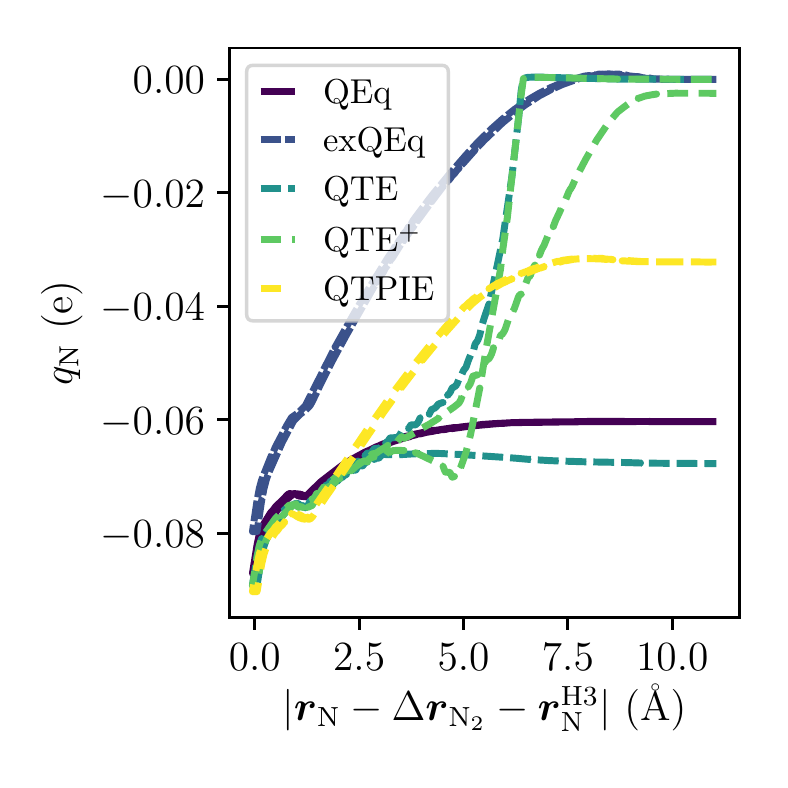}
}
\caption{The charge of the upper nitrogen atom $q_\mathrm{N}$, which is bonded to another, lower nitrogen atom (forming a nitrogen molecule) is displayed against the distance to the hollow adsorption site $\boldsymbol{r}_\mathrm{N}^\mathrm{H3}$ on the AlN($001$) surface.  The $N_2$ bond ($|\Delta \boldsymbol r_\mathrm{N_2}|$ = 1.1 \r A) is parallel to the surface normal.}
\label{fig:qAlNN2_2}
\end{minipage}
\end{figure}

In the following, the charge exchange for the adsorption and desorption of a nitrogen molecule at the hollow  AlN(001) surface site is investigated. The molecule is aligned in parallel to the surface normal and its bond length is kept constant ($|\Delta \boldsymbol r_\mathrm{N_2}|$ = 1.1 \r A). The charge of individual nitrogen atoms is shown in Figure~\ref{fig:qAlNN2_1} and Figure~\ref{fig:qAlNN2_2}. For the surface bonded system, all charge models except exQEq agree with one another. This exception is caused by the tapered electronegativity of the outer nitrogen atom. For an increasing distance between the surface and the molecule, QEq and QTPIE predict an artificially charged molecule (less significant for QTPIE). This phenomenon is also observed during the desorption when applying QTE. By turning QTE into QTE$^+$ this artefact is resolved, so that the distant N$_2$ molecule is charge neutral. This hold also for the exQEq model.

\begin{figure}[t]
\begin{center}
\resizebox{8cm}{!}{
\includegraphics[width=8cm]{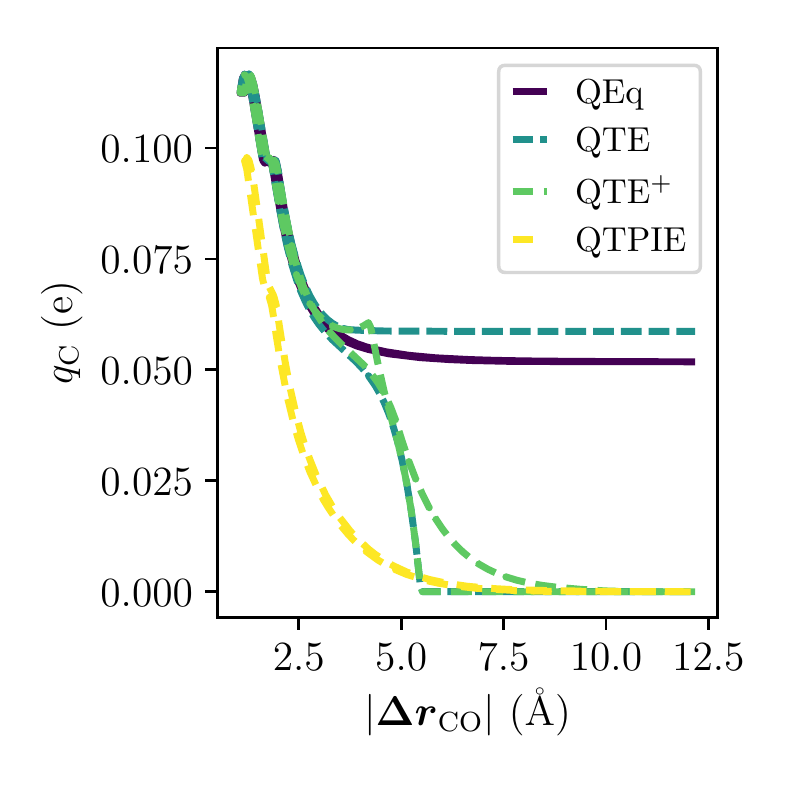}
}
\caption{The charge of a carbon atom $q_\mathrm{C}$ is displayed against the distance to an oxygen atom (forming eventually carbon monoxide).}
\label{fig:qCOC}
\end{center}
\end{figure}

Second, the charge variation of a carbon atom is shown in Figure~\ref{fig:qCOC} as a function of the distance to an oxygen atom, forming eventually carbon monoxide. The charge of the oxygen atom can be easily determined by charge neutrality and therefore equals the negated charge of the carbon atom, $q_\mathrm{O}=-q_\mathrm{C}$. Initially, the two atoms are separated from each other. QEq describes both atoms as charged particles irrespective of distance. Here, this issue cannot be resolved by exQEq, since there is no straightforward way of defining a surface or a reduced charge transfer zone. Once again, QTE describes the separated atoms as neutrals and when decreasing the interatomic distance, agrees more and more with QEq. However, during the dissociation of the molecule, both atoms remain charged. QTE$^+$ overcomes this issue and allows for similar association and dissociation charge states. While QTPIE starts off with neutral atoms, the final charge distribution for carbon monoxide differs from QEq. As generally suggested in the frame of QTPIE, the parameters have to be refitted, but the model describes the correct trend.

\begin{figure}[t]
\begin{minipage}[t]{0.49\linewidth}
\resizebox{8cm}{!}{
\includegraphics[width=8cm]{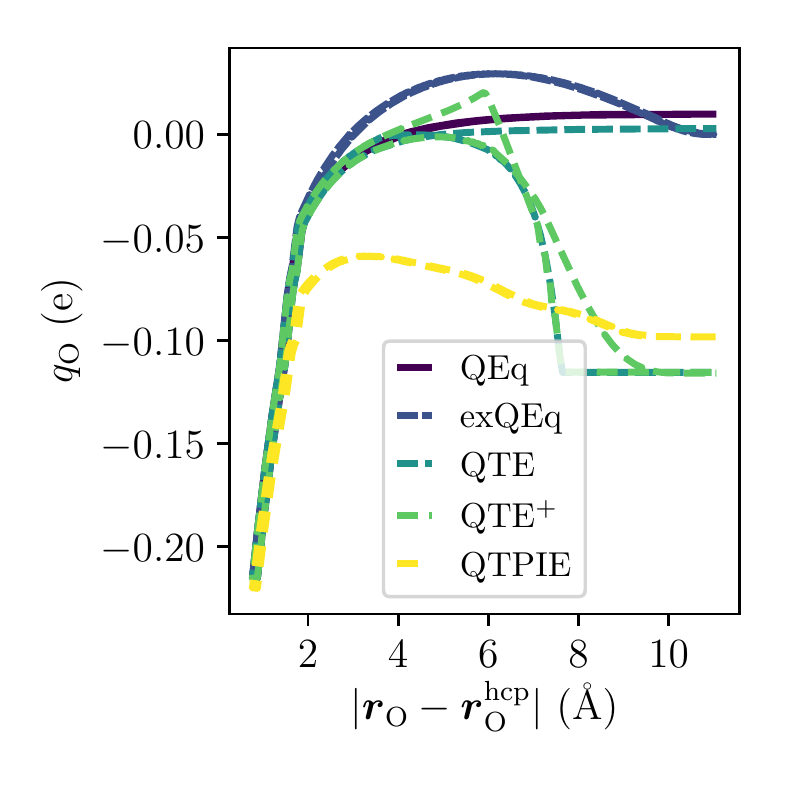}
}
\caption{The charge of an oxygen atom  $q_\mathrm{O}$, which is bonded to a carbon atom (forming carbon monoxide), is displayed against the distance to the hcp adsorption site $\boldsymbol{r}_\mathrm{O}^\mathrm{hcp}$ on the Ti(0001) surface. The CO bond ($r_\mathrm{CO}$ = 1.12 \r A) is parallel to the surface normal.}
\label{fig:qTiCOO}
\end{minipage}
\hfill
\begin{minipage}[t]{0.49\linewidth}
\resizebox{8cm}{!}{
\includegraphics[width=8cm]{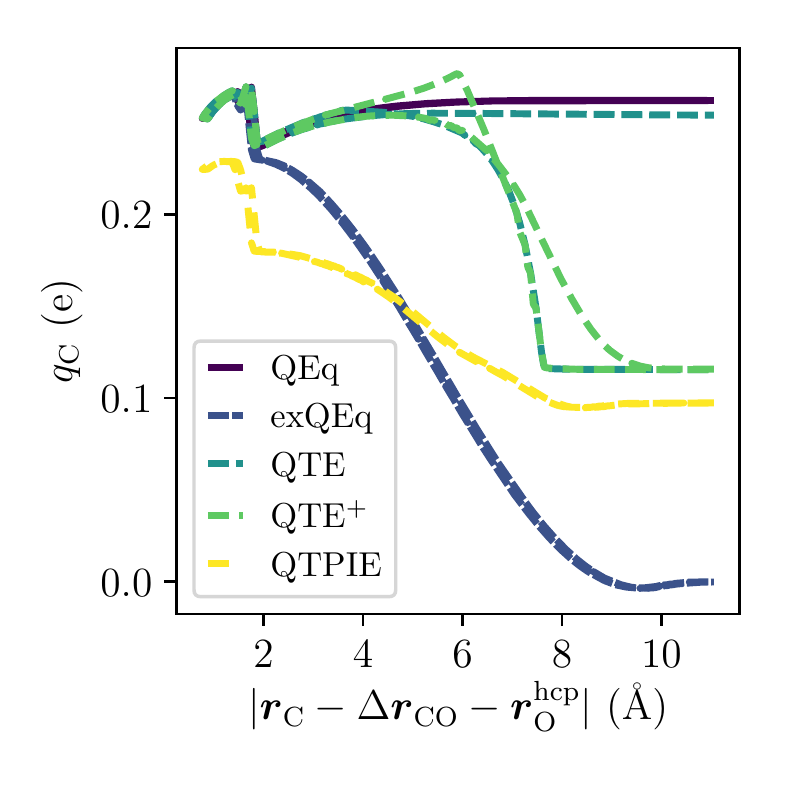}
}
\caption{The charge of a carbon atom  $q_\mathrm{C}$, which is bonded to an oxygen atom (forming carbon monoxide), is displayed against the distance to the respective oxygen hcp adsorption site $\boldsymbol{r}_\mathrm{O}^\mathrm{hcp}$ on the Ti(0001) surface. The CO bond ($r_\mathrm{CO}$ = 1.12 \r A) is parallel to the surface normal.}
\label{fig:qTiCOC}
\end{minipage}
\end{figure}

Next, the charge transfer during the adsorption and desorption of a carbon monoxide molecule at the Ti(0001) hcp surface site is studied. The molecule is aligned in parallel with the surface normal (oxygen atom is closer to the surface) and its bond length is constant ($|\Delta \boldsymbol r_\mathrm{CO}|$ = 1.12 \r A). The charge of the carbon and the oxygen atom is shown in Figure~\ref{fig:qTiCOO} and  \ref{fig:qTiCOC}, respectively. Once again, QEq may be considered as a reference for the charge state at the surface. At the same time, QEq fails to describe the absent charge transfer with the surface for the distant molecule. While exQEq succeeded in repairing this artificial global charge transfer for a single adatoms, the charge distribution in case of molecules are nonphysical: Instead of having a neutral molecule, exQEq describes each individual atom as being neutral. QTE is the only model that is capable of describing the charge states of both subsystems at the same time, while being not in contact with each other. When the molecule is moved closer to the surface, QTE and QEq agree more and more with each other. At the adsorption site, an almost exact agreement is observed. During the desorption process, a net charge remains on the molecule. However, to enforce a charge neutral molecule after its desorption, QTE$^+$ can be applied. While the QTPIE model is able to provide the required trend during this particular surface process, the absolute values differ throughout the distance variation. 

To summarize this section, the QEq model is found to be solely suited for the simulation of surface slabs without a gas phase, i.e., atoms/molecules not bonded to the surface. The QTPIE model extends the applicability to molecules (formation as well as dissociation). However, the simultaneous simulation of molecules or atoms in combination with surfaces is shown to be less reliable and may be avoided. The model also demands reparametrization, which has to be conducted to rigorously test this capability. The exQEq model provides an extension to the QEq model that allows for the description of single atoms impinging or leaving surfaces, but the method is not suited for the coexistence of molecules. The QTE model is found to describe the formation process for any kind of surface interaction appropriately, but fails at the corresponding fragmentation steps. This may be reasoned by possibly inaccurately described ionization energies. However, this issue is resolved by turning QTE into QTE$^+$, enforcing charge neutral subsystems. It is demonstrated that the QTE$^+$ model is well suited for all herein considered scenarios, e.g., adsorption and desorption of atoms and molecules as well as the dissociation and recombination of the latter.

\subsection{External electric fields}
\label{ssec:External elecrtic fields}

\begin{figure}[t]
\begin{minipage}[t]{0.49\linewidth}
\resizebox{8cm}{!}{
\includegraphics[width=8cm]{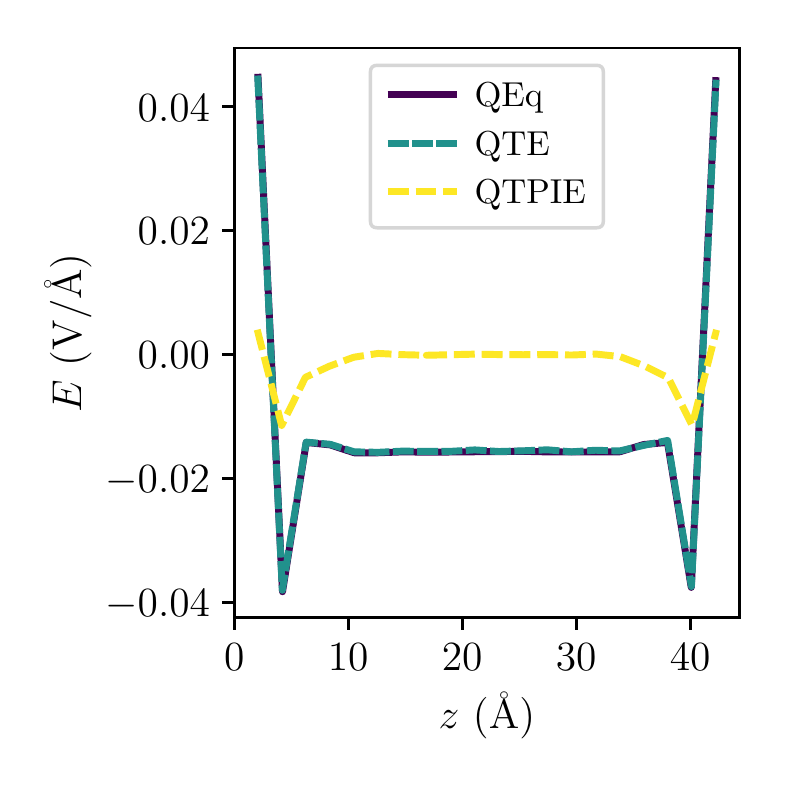}
}
\caption{The average depolarization field along a TiN slab (conductor) is shown for an external electric field 0.01 V/\r A.}
\label{fig:extETiN}
\end{minipage}
\hfill
\begin{minipage}[t]{0.49\linewidth}
\resizebox{8cm}{!}{
\includegraphics[width=8cm]{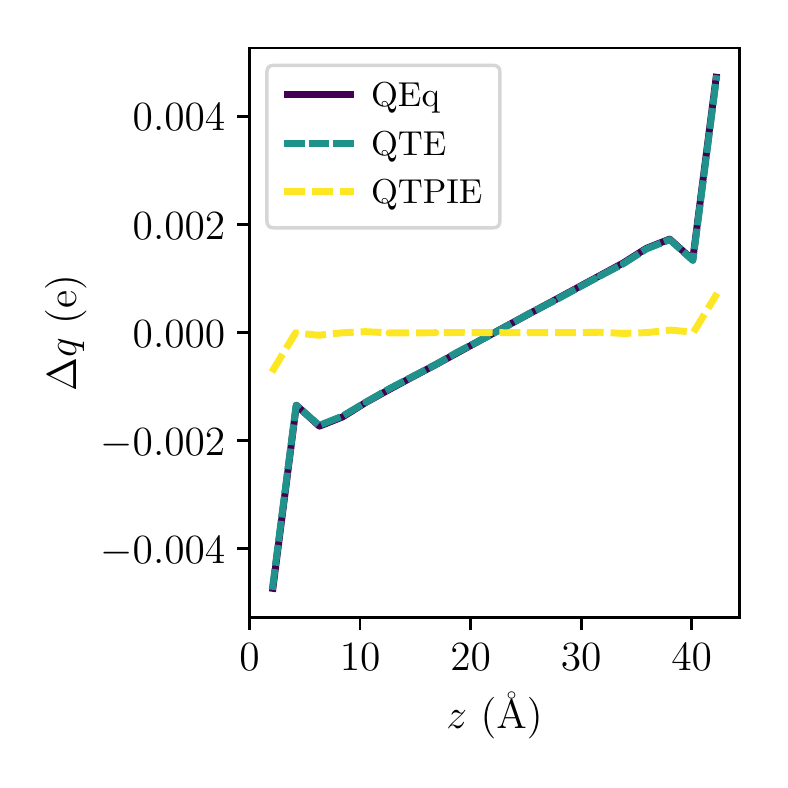}
}
\caption{The average change of the atomic charge distribution along a TiN slab (conductor) induced by the external electric field 0.01 V/\r A.}
\label{fig:delQTiN}
\end{minipage}
\end{figure}

\begin{figure}[t]
\begin{minipage}[t]{0.49\linewidth}
\resizebox{8cm}{!}{
\includegraphics[width=8cm]{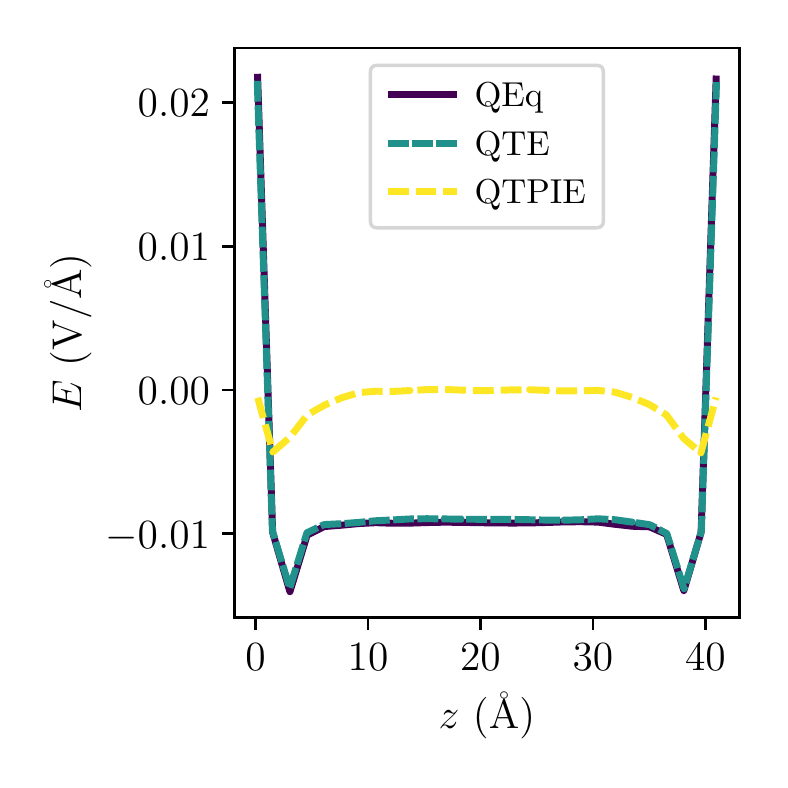}
}
\caption{The average depolarization field along a TiO$_2$ slab (insulator) is shown for an external electric field 0.01 V/\r A.}
\label{fig:extETiO2}
\end{minipage}
\hfill
\begin{minipage}[t]{0.49\linewidth}
\resizebox{8cm}{!}{
\includegraphics[width=8cm]{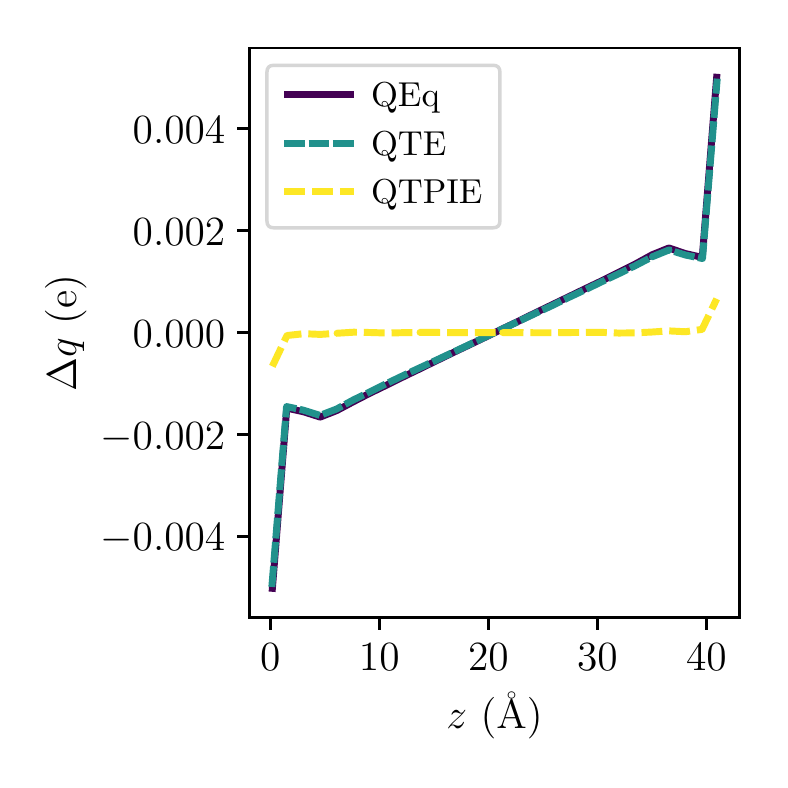}
}
\caption{The average change of the atomic charge distribution along a TiO$_2$ slab (insulator) induced by the external electric field 0.01 V/\r A.}
\label{fig:delQTiO2}
\end{minipage}
\end{figure}

In the following, as example cases, the TiN (conductor) and TiO$_2$ (insulator) polarization due to an externally applied electric field (chosen to be 0.01 V/\r A) are reported for QEq, QTE and QTPIE. The electric field is applied along the [001] direction ($z$-axis). The slabs' thicknesses are chosen to be approximately 40 \r A, which is larger than two times the COMB3 cutoff radius ($2\times11.0$ \r A). The methodology for the evaluation of the depolarization field is described elsewhere \citep{gergs2018integration}. The average depolarization field $E$ for TiN and TiO$_2$ are presented in Figure~\ref{fig:extETiN} and \ref{fig:extETiO2}, respectively. The underlying change of the averaged atomic charge distribution for TiN and TiO$_2$ are presented in Figure~\ref{fig:delQTiN} and \ref{fig:delQTiO2}, respectively. First, it is shown, that none of the particular models is capable of distinguishing between an insulator and a conductor in a self-consistent manner. QEq and QTE describe either system as ideal conductor, fully compensating the externally applied electric field. However, it is important to note, that due to the limitations, which are inherent to the utilization of cutoff radii, polarization charges by means of surface as well as space charges are present. This contradicts the interpretation of an ideal conductor. As expected, QEq and QTE exactly agree with one another. On the other hand, QTPIE leads to a small change of the surface charges as well as negligible compensation of the external electric field. The reduced interaction range due to the cutoff radius limits the model's validity. 

In summary, QEq and QTE may be used to model polarization effects of conductive material systems (e.g., TiN). The application of QTPIE must be limited to non-conductive capacitor arrangements (e.g., metal/TiO$_2$/metal) with externally applied voltages. 

\section{Performance}
\label{sec:performance}

In this section, the CPU time demanded by the fluctuating charge models QEq, QTPIE, QTE and QTE$^+$ are compared with each other for one example system (i.e., AlN002) to provide a first impression of the particular performance. The surface slab consists of $7\times4\times6$ unit cells, whereas the bottom two layers are immobile. The equilibration tolerance for the maximum charge force is set to 0.1 V for any charge model.

\begin{figure}[t]
\begin{minipage}[t]{0.49\linewidth}
\resizebox{8cm}{!}{
\includegraphics[width=8cm]{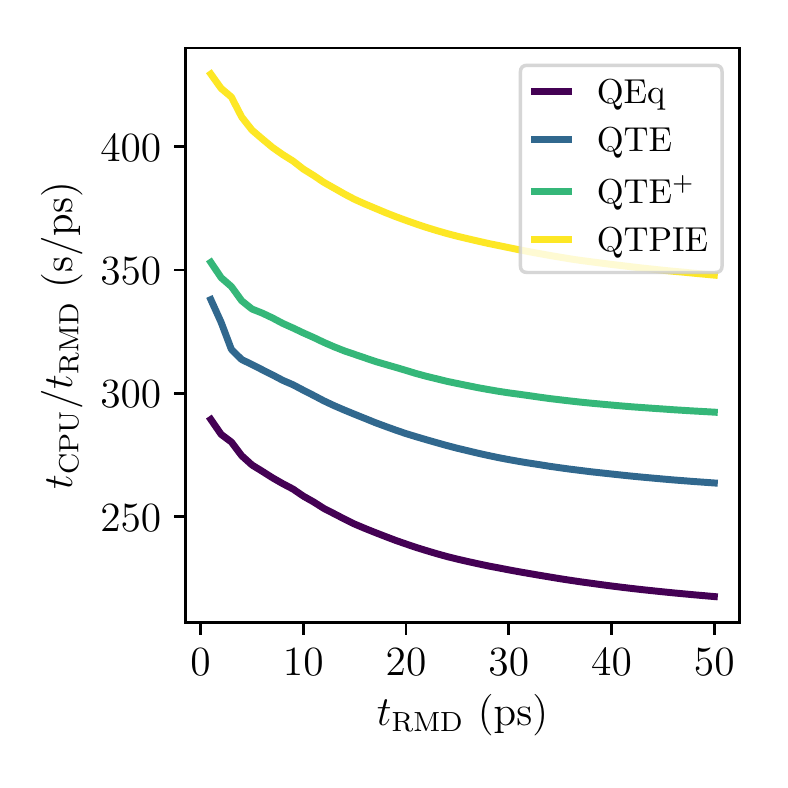}
}
\caption{The average change of the atomic charge distribution along a TiO$_2$ slab (insulator) induced by the external electric field 0.01 V/\r A.}
\label{fig:performance_rMD}
\end{minipage}
\hfill
\begin{minipage}[t]{0.49\linewidth}
\resizebox{8cm}{!}{
\includegraphics[width=8cm]{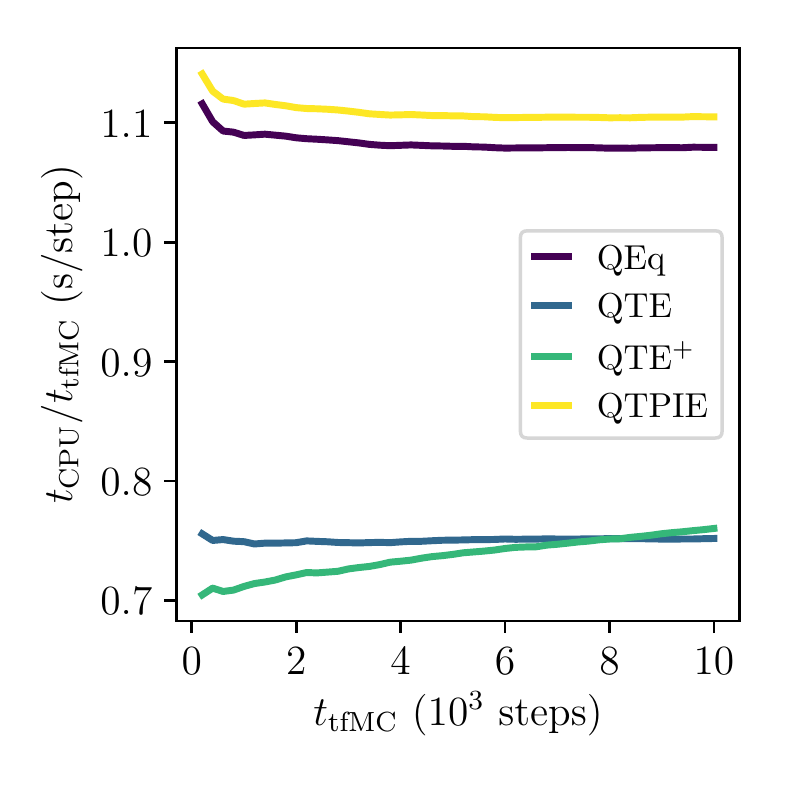}
}
\caption{The average change of the atomic charge distribution along a TiO$_2$ slab (insulator) induced by the external electric field 0.01 V/\r A.}
\label{fig:performance_tfMC}
\end{minipage}
\end{figure}

First, the canonical ensemble is simulated over a period of 50 ps ($dt=0.2$ fs) by utilizing a Nose-Hoover thermostat (300 K with a damping constant of 100 fs), as implemented in LAMMPS \citep{plimpton1995fast, martyna1994constant, parrinello1981polymorphic, tuckerman2006liouville, shinoda2004rapid, dullweber1997symplectic}. The required CPU time $t_\mathrm{CPU}$ per RMD time $t_\mathrm{RMD}$ is shown in Figure~\ref{fig:performance_rMD}. For the provided case, QEq outperforms QTE, QTE$^+$ and QTPIE by factors of 1.21, 1.34, 1.60, respectively. This is due to the additional computations of the overlap integrals, inter processor communication as well as nested for-loops. Due to the chosen precision and step size, typically only one execution for any charge model is sufficient to achieve convergence (neglecting the initial charge equilibration).

Second, the  canonical ensemble is simulated for $10^4$ steps utilizing the time-stamped force-bias Monte Carlo (tfMC) algorithm (300 K with a maximal displacement length of $\Delta = 0.19$ \r A), as implemented in LAMMPS \citep{plimpton1995fast, bal2014time, mees2012uniform, neyts2014combining}. The required CPU time $t_\mathrm{CPU}$ per tfMC time step $t_\mathrm{tfMC}$ is shown in Figure~\ref{fig:performance_tfMC}. For the present case, QTE and QTE$^+$ outperform QEq and QTPIE by a factors of 1.44. The tfMC method inherently leads to larger distortions and consequently requires more iterations of the charge equilibration schemes to meet convergence. While on average QEq and QTPIE demand 72.6 iterations per step, QTE and QTE$^+$ on average require only 43.4 iterations per step. The computational resources  required per individual iteration of the two latter may be higher in comparison to QEq, but due to smaller number of total iterations, QTE and QTE$^+$ are more efficient than QEq.

To summarize, while the computational cost of all charge models is of same order of magnitude, the fastest model evaluation appears to be case dependent.


\section{Conclusion}
\label{sec:discussion}

When studying surfaces, the charge equilibraiton procedures of EEM, QEq and ES+ are only sufficient in cases without association, dissociation, adsorption, desorption, sputtering, among others. This is reasoned in the underlying global charge transfer, as discussed in Section~\ref{ssec:extended_lagrangian_method_for_EEM_QEq_and_ES+} and demonstrated in Section~\ref{ssec:Surface processes}. This limitless charge transfer is addressed by the extension, provided in \ref{ssec:extended_lagrangian_method_for_EEM_QEq_and_ES+}. This enables adsorption as well as desorption simulations of single atoms approaching or leaving the surface. However, molecules are insufficiently described, i.e. the intramolecular charge exchange is weakened and eventually omitted. Without this extension, any system is described as ideal conductor (metal), which potentially becomes a challenging as assumption when insulating material systems are studied under the effect of externally applied electric fields.

Specifically for metal/insulator/metal arrangements with applied voltages, QTPIE may be used to describe the depolarization field (e.g., in the frame of memristive mechanisms). It is important to note, though, that polarization is only modeled along the interatomic bonds. This is an inherent limitation for any atomic charge model. For a more sophisticated polarization model, as realized for certain cases in COMB3, dipoles as additional degree of freedom have to be introduced \citep{stern1999fluctuating, liang2013classical}. One also has to take care, that the corresponding change of the atomic charge distribution does not deviate too much from its equilibrium state. Due to the underlying second-order Taylor expansion, fluctuating charge models are only valid for a particular finite range. However, in case of surface simulations, as shown in Section~\ref{ssec:Surface processes}, QTPIE is well suited for the description of molecules, but is less reliable when the charge exchange between surfaces and distant particles is studied. Unfortunately, QTPIE also requires any set of EEM, QEq and ES+ parameters to be refitted.

The charge transfer equilibration (QTE) model described in Section~\ref{ssec:charge_transfer_equilibration} is a generalization of QEq (or EEM, ES+) and corresponds to the latter for a hypothetical global charge transfer, $S_{ij}=1$. As depicted in Section~\ref{ssec:Surface processes}, for bonded systems, QTE tends to agree almost exactly with QEq. Hence, when switching from QEq to QTE, the polarizability is not altered. Though, the range and intensity of the charge transfer is self-consistent. The orbital overlap distribution of the respective atom geometry can therefore be interpreted as a network for the interatomic charge transfer. Hence, QTE is theoretically suited for studying any kind of surface process. In practical combination with COMB3, however, systems split up in charged subsystems. This phenomenon can be addressed by turning QTE to QTE$^+$, enforcing charge neutral subsystems. However, this necessity should be revisited when another RMD potential is utilized. As elaborated in Section~\ref{ssec:Surface processes}, QTE$^+$ is the only fluctuating charge model considered that is capable of describing the charge transport during all kinds of surface interactions.

While the orbital of the valence electrons can be described by a variety of orbital types, we believe that it should be chosen consistent with the outer RMD potential model. For example, if one would like to replace QEq with QTE or QTE$^+$, an $n$s Slater type orbital (STO) may be used. In combination with COMB3, where a $Z$+$1$s STO is used for the electrostatic interactions (ES+), a $1$s STO should be used for the QTE or QTE$^+$ model. $Z$ is the charge of the nuclei. For the originally published and still in LAMMPS implemented ReaxFF method, where EEM is applied and thus, no orbitals are used, a linear combination of three Gaussian type orbitals (STO-3G) could be considered \citep{van2001reaxff}. The ACKS2 model is utilized in the most recent revision of the ReaxFF method, which, however, needs yet to be implemented in LAMMPS \citep{senftle2016reaxff}. For any interaction potential, the orbital exponent for the QTE model may either be taken from the particular model or be fitted to the respective overlap integrals.

The mirror boundary condition (MBC), described in Section~\ref{sec:mirror_boundary_condition}, speeds up RMD surface simulations that employ variable charge models. Our experience with the MBC indicates that RMD (i.e., COMB3) simulations of thin ($\approx$ 30 \r A $\times$ 30 \r A $\times$ 30 \r A) surface slabs can be sped up by a factor of approximately 2. The reduced computational cost may, however, vary as a function of potential choice as well as system size. As pointed out in Section~\ref{sec:implementation}, there are a few exceptions that have to be considered when implementing the MBC.

\section*{Acknowledgement}

Funded by the Deutsche Forschungsgemeinschaft (DFG, German Research Foundation) – Project-ID 138690629 (TRR 87) and Project-ID 434434223 (SFB 1461).

\section*{ORCID iDs}
T. Gergs: \url{https://orcid.org/0000-0001-5041-2941}\\
F. Schmidt: \url{https://orcid.org/0000-0001-6623-0464} \\
T. Mussenbrock: \url{https://orcid.org/0000-0001-6445-4990} \\
J. Trieschmann: \url{https://orcid.org/0000-0001-9136-8019}


%

\end{document}